\documentstyle[aps,preprint,epsfig,tighten,amssymb]{revtex}
\begin{document}
\draft


\title{
Probing the strange quark condensate by di-electrons 
from $\phi$ meson decays
in heavy-ion collisions at SIS energies 
}


\author{
B. K\"ampfer$^a$,
O.P. Pavlenko$^{a,b}$,
S. Zschocke$^a$
}

\address{
$^a$Forschungszentrum Rossendorf, PF 510119, 01314 Dresden,
Germany\\
$^b$Institute of Theoretical Physics,
03143 Kiev, - 143, Ukraine\\
}

\maketitle

\begin{abstract}
QCD sum rules predict that
the change of the strange quark condensate $\langle \bar s s \rangle$
in hadron matter at finite baryon density causes a shift of the
peak position of the
di-electron spectra from $\phi$ meson decays.
Due to the expansion of hadron matter in heavy-ion
collisions, the $\phi$ peak suffers a smearing governed by the
interval of density in the expanding fireball, which appears as
effective broadening of the di-electron spectrum in the
$\phi$ region.
The emerging broadening is sensitive to the
in-medium change of $\langle \bar s s \rangle$. This allows
to probe directly in-medium modifications of $\langle \bar s s \rangle$
via di-electron spectra in heavy-ion collisions at SIS energies
with HADES.
\\[6mm]
PACS number(s): 25.75.+r, 14.60.-z, 14.60.Cd
\end{abstract}

\section{Introduction} 

The starting heavy-ion collision experiments with the 
{\bf H}igh {\bf A}cceptance {\bf D}i{\bf E}lectron {\bf S}pectrometer
(HADES) \cite{HADES} at the heavy-ion synchrotron SIS at GSI/Darmstadt
provide the rare opportunity to measure via the di-electron channel 
($e^+ e^-$)
in-medium modifications of the light vector mesons with high accuracy.
High accuracy is particularly important for the $\phi$ meson since the 
commonly
expected change of the $\phi$ meson spectral function in a nuclear medium
is not too strong.

From the theoretical side there are at least two motivations 
for studying in detail
the in-medium modifications of the $\phi$ meson. First,
as shown within various
QCD sum rule approaches \cite{Hatsuda,our} the mass shift of the $\phi$ meson
in nuclear matter is directly related to the in-medium change of the strange
quark condensate $\langle \bar s s \rangle$ and, therefore, provides a direct
access to the chiral symmetry restoration in the strange sector
\cite{BR_2002}. This is in contrast to the
$\rho$ and $\omega$ mesons whose in-medium mass shifts are sensitive
to both chiral condensate $\langle \bar q q \rangle$ and 
the poorly known four-quark condensate $\langle (\bar q q)^2 \rangle$,
in particular for the $\omega$ meson  
\cite{our}. (For the $\rho$ meson there is also a strong broadening effect
expected \cite{Weise,Rapp_Wambach}, so that the $\rho$ meson can even disappear
as a quasi-particle in a dense nucleon medium.)
The second reason to focus on the $\phi$ meson in-medium effects is related
to the great interest in measuring and understanding the strangeness content
of the nucleon and nuclear matter \cite{Strange_matter_workshop}. This issue has a lot
of physical applications in various branches such as deep-inelastic
lepton-nucleon scattering, production of strange particles in hadron and
nuclear collisions, speculations on the nature of dark matter etc.

Within model approaches based on effective hadronic Lagrangians 
the $\phi$ meson mass in a nuclear medium is affected by the 
$\phi N$ scattering.
Because of the small real part of the $\phi N$ scattering amplitude 
the corresponding
change of the $\phi$ meson mass appears to be tiny in nuclear matter
at saturation density. The $\phi$ meson mass shift remains still small
in such approaches even if one takes into account various effects related
to the $K \bar K$ in-medium loop in the $\phi$ meson self-energy
\cite{Weise,Cabrera}. This means that the main cause of a possible $\phi$ meson
mass shift in a nuclear medium is connected with the change of the density
dependent  $\langle \bar s s \rangle$ condensate which actually acts here as
a density dependent mean field \cite{Hatsuda,BR_2002,Asakawa_Ko}. 
The most consistent way
to incorporate mean field effects is through the QCD sum rules, 
which match directly
the in-medium spectral function with the density dependent 
strange quark condensate.

In the present work we employ the results \cite{Hatsuda,our} 
of the QCD sum rule evaluations of the
$\phi$ meson mass shift at finite baryon density to find the
possible modifications of the di-electron spectrum in the region of the $\phi$ peak
in heavy-ion collisions at SIS energies. Due to the collective expansion of the matter
formed in the course of heavy-ion collisions the baryon density $n$ depends on time
so that a time average of the mass shift is expected. As a result, the mass shift 
is smeared out over an interval related to the temporal change of the density
and looks like
an ''effective'' broadening of the $\phi$ peak. It is remarkable that this effective
$\phi$ broadening reflects essentially the in-medium behavior of the strange
condensate $\langle \bar s s \rangle$ but is almost insensitive to the collision 
broadening. This offers a chance to probe the in-medium strange quark condensate via 
measuring the effective broadening of the $\phi$ peak in the $e^+ e^-$ spectrum.
          
To get the di-electron spectrum from the $\phi$ meson decays we use a transparent
hydrodynamical model for the space-time evolution of the matter including the
obvious possibility that the $\phi$ meson does not chemically equilibrate.
In spite of the schematic character of our dynamical model, 
it delivers results which are
confirmed by the transport model calculations of BUU type \cite{Wolf}.

\section{Di-electron spectra from in-medium $\phi$ decays}

Within the linear density approximation and for not too high temperatures
$T < 100$ MeV the strange quark condensate in hadronic matter can be written as
\begin{equation}
\langle \bar s s \rangle_{\rm matter} = \langle \bar s s \rangle_0
+ \sum_h \frac{\langle h \vert \bar s s \vert h \rangle}{2 M_h} \, n_h,
\end{equation} 
where $\langle \bar s s \rangle_0$ is the vacuum condensate,
$\langle h \vert \bar s s \vert h \rangle$ denotes the matrix element corresponding
a one-hadron state (we employ the normalization $\langle h \vert h \rangle = 2 E_h$
with $E_h \approx M_h$ with $E_h$ and $M_h$ as respective energy and mass of
the hadron species $h$), $n_h$ stands for the hadron density, and the sum
runs over all hadrons in the medium, i.e., $h = N, \Delta, \Sigma, K, \cdots$.
For the nucleon matrix element $\langle N \vert \bar s s \vert N \rangle$
the dimensionless parameter $y$ is widely used to specify the strangeness
content in the nucleon via
\begin{equation}
\frac{\langle N \vert \bar s s \vert N \rangle}{2 M_N} = y \frac{\sigma_N}{2 m_q},
\end{equation} 
where $\sigma_N$ is the nucleon sigma term and $m_q$ the light quark mass
$m_q = \frac12 (m_u + m_d)$. Since the strangeness content of the nucleon is a 
yet poorly known quantity and matter of debate so far (see, for instance,
the QCD lattice calculations in \cite{Liu,Michael} and the phenomenological
study in \cite{Alberico})
we shall vary below $y$ in the interval $0 \cdots 0.2$ to get the corresponding
modifications of the di-electron spectra. We also simplify eq.~(1) by the 
replacement 
$ \sum_h \frac{\langle h \vert \bar s s \vert h \rangle}{2 M_h} \, n_h
\to y \frac{\sigma_N}{2 m_q} \, n_N$ keeping in mind
that the presence of other hadron states in the sum can only increase the
value of $y$ \cite{Asakawa_Ko}, so that our choice reflects a lower limit
of the strangeness content of hadronic matter.

Basing on the above parameterization of the strange quark condensate in hadron
matter one can perform the QCD sum rule evaluations which give for the $\phi$
meson mass the density dependence in leading order \cite{our}   
\begin{equation}
m_\phi = m_\phi^{\rm vac} (1 - 0.14 y \frac{n_N}{n_0}),
\end{equation}
where $m_\phi^{\rm vac}$ denotes the vacuum value of the $\phi$ mass,
and the nuclear matter saturation density is $n_0 = 0.15$ fm$^{-3}$.

The di-electron rate from the in-medium $\phi$ meson decays 
in ideal gas approximation at temperature $T$ is given by 
\begin{equation}
\frac{dN}{d^4 x \, d^4 Q} = \frac{6}{(2 \pi)^3} 
\exp \left( - \frac{u \cdot Q}{T} \right)
M \, \Gamma_{\phi \to e^+ e^-} \, A(M^2, m_\phi, \Gamma_\phi^{\rm tot}) 
\end{equation} 
where $\Gamma_{\phi \to e^+ e^-}$ and $\Gamma_\phi^{\rm tot}$ are the di-electron
and total decay widths of the $\phi$ meson, $Q_\mu$ denotes the four-momentum
of the di-electron with invariant mass $M^2 = Q^2$, and $u_\mu$ is the four-velocity
of the emitting medium. 
We use the Breit-Wigner parameterization of the spectral function  
\begin{equation}
A (m, m_\phi, \Gamma_\phi^{\rm tot})
= \frac1\pi \frac{m_\phi \Gamma_\phi^{\rm tot}}
{(m^2 - m_\phi^2)^2 + (m_\phi \Gamma_\phi^{\rm tot})^2},
\end{equation}
having in mind that the $\phi$ meson may survive as single-peaked quasi-particle
excitation \cite{Weise}, while in-medium the width is expected to be noticeably
enlarged, at least by collision broadening to 
$\Gamma_\phi^{\rm tot} = 20 \cdots 30$ MeV. Due to four-momentum conservation
in the direct decay $\phi \to e^+ e^-$ one also gets the relation $m^2 = M^2$
interrelating eqs.~(4, 5). 

To obtain the di-electron spectrum from the rate (4) one needs to specify the space-time
evolution of the matter. According our experience \cite{Gallmeister} the detailed
knowledge of the space-time evolution is not necessary to describe the experimental
di-electron spectra in relativistic heavy-ion collisions in the low-mass region
\cite{Agakhiev}. In this line, and also to avoid too many parameters, we employ
here a variant of the blast wave model (cf.\ \cite{our3}) with a constant
radial expansion velocity $v_r = 0.3$. We take for the initial baryon density
$n_N (t = 0) = 3 n_0$, as suggested by transport code simulations for the maximum
density at SIS-18 (cf.\ \cite{HW}), and $n_N(t_{\rm f.o.}) = 0.3 n_0$ for the
freeze-out density. The evolution of the matter, i.e., $n(t)$,
is defined by the baryon conservation
within the fireball volume; we use as baryon participant number $N_B = 330$, and
$T(t = 0) = 90$ MeV as initial temperature of the systems. These numbers are to
be understood as spatial averages, i.e., gradients are not explicitly taken into
account. The freeze-out temperature $T_{\rm f.o.} = 50$ MeV, deduced
in \cite{Cleymans} from hadro-chemistry, 
and the initial temperature are linked by entropy
conservation. A convenient parameterization for the relevant  
section of the isentrope with
specific entropy 5 per baryon is, e.g., $T(t) = (a + b n(t))^{1/6}$ with $a$ and $b$
given by the initial and freeze-out conditions.

Within the described fireball dynamics we consider two cases for the evolution
of the $\phi$ multiplicity:\\
(i) the number of $\phi$ mesons is assumed to be governed by chemical equilibrium,
so that it is proportional to $\exp ( - m_\phi / T )$, and\\
(ii) the $\phi$ mesons are still in thermal equilibrium with the bulk of matter
but do not maintain chemical equilibrium; to be specific, we take the number
of as constant thus assuming a weak $\phi$ inelasticity. This assumption
is supported by BUU transport calculations \cite{HW}.

After freeze-out, the $\phi$ meson decays in vacuum contribute to the
di-electron spectrum according to
\begin{equation}
\frac{d N}{d M^2} = Br_{\phi \to e^+ e^-} \,
N_\phi(T_{\rm f.o.}, M^2) \,
A(M^2, m_\phi^{\rm vac}, \Gamma_\phi^{\rm tot, vac}),
\end{equation} 
where $Br_{\phi \to e^+ e^-}$ is the branching ratio of the $\phi$ decay
$\phi \to e^+ e^-$ in vacuum, 
and $N_\phi(T_{\rm f.o.}, m_\phi)$ is the number of of $\phi$ mesons at freeze-out.
This number is given in the spherical expansion model for case (i) by
\begin{eqnarray}
N_\phi(T_{\rm f.o}, m_\phi) & = & 
\frac{1}{(2 \pi)^3} \frac{4 \pi R_{\rm f.o.}^3}{\gamma}
\int d p_\perp^2 \pi \sqrt{\frac{2 \pi T_{\rm f.o.}}{\gamma m_\perp}}
\exp \left( - \frac{\gamma m_\perp}{ T_{\rm f.o.}} \right) \times \nonumber \\
& & \left[ \frac{\sinh a_{\rm f.o.}}{a_{\rm f.o.}} (\gamma m_\perp + T_{\rm f.o.}) 
- T_{\rm f.o.} \cosh a_{\rm f.o.} \right]
\end{eqnarray} 
where $m_\perp = \sqrt{m_\phi^2 + p_\perp^2}$, 
$a_{\rm f.o.} = \gamma v_r p_\perp / T_{\rm f.o.}$
and $p_\perp$ is the transverse momentum of the $\phi$ meson;
$\gamma$ is the Lorentz factor related to the transverse expansion
velocity $v_r$.

The number of $\phi$ mesons in case (ii) can be parameterized by including
in eq.~(7) an off-equilibrium chemical potential which is governed by the
condition $N_\phi = const$ during the expansion. 

Note that, at present, the understanding of the $\phi$ dynamics
in heavy-ion collisions at threshold energies is still in the
infancy. The data base is poor \cite{Kotte}, and the interpretation
by transport models meets uncertainties \cite{HW,bk}.
While the multiplicities of other hadrons, after freeze-out, can
fairly well be described by a statistical (equilibrium) model
\cite{Cleymans}, the situation for the $\phi$ mesons remains
unsettled. We are, therefore, left with the two above extreme scenarios.

\section{Results}

The results of our calculations of the di-electron invariant mass spectrum
from $\phi$ decays are exhibited in fig.~1 for the case (i), i.e., where the
$\phi$ meson is assumed to be in chemical equilibrium.
As seen here the in-medium shift of the strange quark condensate, specified by
the parameter $y$, causes a noticeable shift of the peak position due to contributions
of $\phi$ decays inside the medium. One can even expect a double peak
structure \cite{Asakawa_Ko2} which emerges from the superposition of vacuum
and in-medium contributions. Due to the smearing of the $\phi$ meson mass shift
over an interval of density of the expanding fireball, one can actually observe
an effective broadening of the second ''shifted'' peak. Such an effective broadening
increases almost linearly with the parameter $y$ independently of the value
of the $\phi$ meson width $\Gamma_\phi^{\rm tot}$:
As displayed in fig.~2, the linear dependence of the effective broadening on the 
strange quark condensate still holds if one varies the width $\Gamma_\phi^{\rm tot}$
from the vacuum value up to the sizeable value of 20 MeV caused by $\phi N$
collisions and predicted by effective hadronic interaction Lagrangians \cite{Weise}.
This offers a chance to probe the strange quark condensate in hadronic matter
(and thus the parameter $y$) via measuring the effective broadening of the second
''shifted'' peak in the $\phi$ meson region of the di-electron spectrum.

Due to the assumed chemical equilibrium, the number of $\phi$ mesons
drops rapidly with decreasing temperature. Therefore, the di-electron
signal from $\phi$ mesons decaying after freeze-out appears rather
small, in contrast to naive expectations assuming a ''life time of 45 fm/c''. 
The peak height of the vacuum decay contribution varies strongly with
changing values of $T_{\rm f.o.}$.

In fig.~3 we show the in-medium modifications of the di-electron spectrum from
$\phi$ mesons in the case of chemical off-equilibrium with fixed number of
$\phi$ mesons during the evolution. Here the in-medium modification
of the strange quark condensate is reflected by the appearance of the characteristic
l.h.s.\ shoulder additionally to the vacuum $\phi$ peak. The effective broadening
of the shoulder still increases linearly with the parameter $y$ and is 
insensitive to variations of the width $\Gamma_\phi^{\rm tot}$
within the given range.
In contrast to the above assumed chemical equilibrium situation, the
vacuum decay contribution is here much stronger. This is since the 
$\phi$ multiplicity is kept constant, according to our scenario (ii)
and in line with \cite{HW}. Correspondingly, also the in-medium decay
is somewhat stronger than in case (i). Nevertheless, the separation
of this in-medium $\phi$ signal will be an experimental challenge.

We have here displayed only the di-electron spectra from exclusive
decays $\phi \to e^+ e^-$. Actually, there will be contributions
from $\phi$ Dalitz decays which give additional strength, sufficiently
far at the low-mass side of the direct $\phi$ signal. 
Both contributions sit on a 
steeply falling background which emerges essentially from the
high-mass $\rho$ decays, bremsstrahlung etc. 
As shown in \cite{our3}, the $\phi$ signal 
may be clearly separated from this background. For further studies
of the feasibility of identifying the $\phi$ signal within the
HADES acceptance we refer the interested reader to \cite{Schon}.

\section{Summary}

In summary, the expansion of nuclear matter after
the maximum compression stage in the course of heavy-ion collisions
is expected to
cause a broadening of the di-electron spectrum in the $\phi$ meson region.
The amount of the broadening appears to be almost insensitive
to the collision broadening in the nuclear medium. 
Rather, the broadening is directly related
to the dependence of the strange quark condensate $\langle \bar s s \rangle$
on the nucleon density which, in turn, is related to the strangeness
content of the nucleon. 
This gives a good opportunity to probe the corresponding in-medium
modification of $\langle \bar s s \rangle$ 
with HADES in heavy-ion collisions. The impact parameter dependence,
which could be used to study details of the density dependence of
$\langle \bar s s \rangle$, deserves further investigations.
Supplementary information can be gained from the di-electron
spectrum in the $\phi$ region in reactions of elementary projectiles
with nuclei. 

\subsubsection*{Acknowledgments}

Useful discussions with H.W. Barz, Gy. Wolf and G. Zinovjev are
acknowledged. The work is supported by BMBF grant 06DR921.
O.P.P. thanks for the warm hospitality  of the hadron theory group
in the Research Center Rossendorf/Dresden and the support by
STCU 15a, CERN-INTAS 2000-349 and NATO-2000-PST.CLG 977482.

\begin{figure}
\begin{center}
\includegraphics[width=7.0cm,angle=-0]{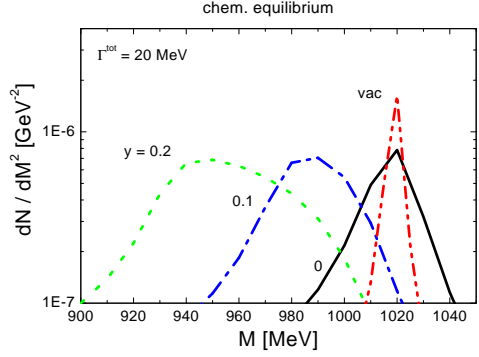}
\caption{Di-electron spectra from $\phi$ meson decays as a function of 
invariant mass for case (i), i.e., assumed chemical equilibrium.
''vac'' labels the decay contribution after freeze-out, while the
curves labeled by $y = 0,$ 0.1 and 0.2 depict the in-medium decay
contributions for various values of the strangeness content of the
nucleon. The total width of the $\phi$ meson is assumed to be
$\Gamma_\phi^{\rm tot} = 20$ MeV.} 
\label{fig_1}
\end{center}
\end{figure}

\begin{figure}
\begin{center}
\includegraphics[width=7.0cm,angle=-0]{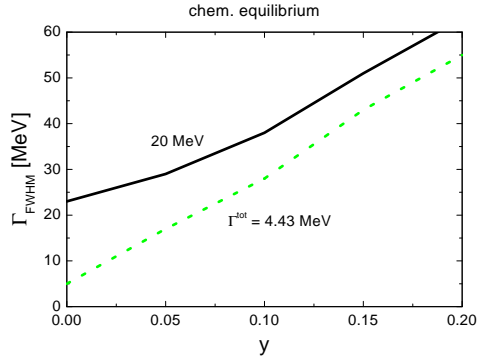}
\caption{The full width half maximum of the in-medium $\phi$ peak 
in the $e^+ e^-$ mass spectrum as a function
of the strangeness content $y$ of nuclear matter
for two values of the total $\phi$ meson decay width 
$\Gamma_\phi^{\rm tot}$.} 
\label{fig_2}
\end{center}
\end{figure}

\begin{figure}
\begin{center}
\includegraphics[width=7.0cm,angle=-0]{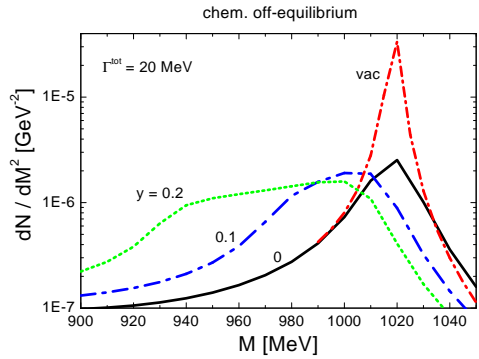}
\caption{As in fig.~1 but for case (ii), i.e., chemical off-equilibrium
of $\phi$ mesons.} 
\label{fig_3}
\end{center}
\end{figure}

\end{document}